\documentclass[reprint,
superscriptaddress,
 amsmath,amssymb,
 aps,
]{revtex4-2}

\usepackage{graphicx}
\usepackage{dcolumn}
\usepackage{bm}

\begin{document}

\preprint{APS/123-QED}

\title{Aluminum Cayley trees as scalable, broadband, multi-resonant optical antennas}

\author{Thomas Simon}
\affiliation{Light, nanomaterials, nanotechnologies (L2n), Universit\'e de Technologie de Troyes, CNRS EMR 7004, 12 rue Marie Curie, Troyes 10000, France}
\author{Xiaoyan Li}
\affiliation{Laboratoire de Physique des Solides, Batiment 510, UMR CNRS 8502, Universit\'e Paris Saclay, Orsay 91400, France}
\author{Jérôme Martin}
\author{Dmitry Khlopin}
\affiliation{Light, nanomaterials, nanotechnologies (L2n), Universit\'e de Technologie de Troyes, CNRS EMR 7004, 12 rue Marie Curie, Troyes 10000, France}
\author{Odile St\'ephan} 
\affiliation{Laboratoire de Physique des Solides, Batiment 510, UMR CNRS 8502, Universit\'e Paris Saclay, Orsay 91400, France}
\author{Mathieu Kociak} 
\affiliation{Laboratoire de Physique des Solides, Batiment 510, UMR CNRS 8502, Universit\'e Paris Saclay, Orsay 91400, France}
\author{Davy G\'erard}
\affiliation{Light, nanomaterials, nanotechnologies (L2n), Universit\'e de Technologie de Troyes, CNRS EMR 7004, 12 rue Marie Curie, Troyes 10000, France}
\email{davy.gerard@utt.fr}


\begin{abstract}
An optical antenna can convert a propagative optical radiation into a localized excitation, and reciprocally. Although optical antennas can be readily created using resonant nanoparticles (metallic or dielectric) as elementary building blocks, the realization of antennas sustaining multiple resonances over a broad range of frequencies remains a challenging task. Here, we use aluminum self-similar, fractal-like structures as broadband optical antennas. Using electron energy loss spectroscopy, we experimentally evidence that a single aluminum Cayley tree, a simple self-similar structure, sustains multiple plasmonic resonances. The spectral position of these resonances is scalable over a broad spectral range spanning two decades, from ultraviolet to mid-infrared. Such multi-resonant structures are highly desirable for applications ranging from non-linear optics to light harvesting and photodetection, as well as surface-enhanced infrared absorption spectroscopy.
\end{abstract}

\maketitle

Optical antennas are nanostructures that can convert a freely propagating radiation into a localized excitation and vice-versa \cite{Novotny:2011ko}. As such, they are key elements for any nanophotonic device, acting as entrance and exit ports, energy harvesting systems, directional emitters, or substrates for enhanced spectroscopies (see, e.g., Ref. \cite{agio2013optical} for an overview of the topic). Simple optical antennas can be realized using either a single metallic (plasmonic) nanoparticle forming a monopole antenna \cite{taminiau2008optical} or by two coupled metallic nanoparticles forming the so-called gap antenna, or dipole antenna \cite{muehlschlegel2005resonant}. Illuminated at its surface plasmon resonance frequency, a plasmonic optical antenna yields intense and confined electromagnetic fields in its vicinity \cite{kinkhabwala2009large} and, reciprocally, it scatters and redirect light emitted by an optical source located nearby \cite{taminiau2008optical}. These antennas are however limited to a single operating frequency. Alternatively, optical antennas can be realized with high refractive index dielectric particles sustaining Mie resonances \cite{koshelev2020dielectric}. Compared with their plasmonic counterparts, dielectric antennas sustain both electric and magnetic resonances \cite{kuznetsov2012magnetic}, allowing interaction with magnetic transition dipole moments \cite{bidault2019dielectric}. Dielectric antennas also have low Ohmic losses, but they do not exhibit the level of electromagnetic field enhancement and confinement that plasmonic antennas can reach. 

Multi-resonant plasmonic optical antennas would be particularly useful, notably for non-linear optics and multiple harmonics generation; the antenna could hence be designed to be resonant at the pump wavelength and at the different harmonics wavelengths. Such an approach should lead to extremely efficient wavelength conversion \cite{huttunen2019}. The design of an efficient multi-resonant plasmonic optical antenna requires two main ingredients: (i) a material sustaining surface plasmon resonances over a broad range of wavelengths ; and (ii) a design allowing to tune under control the position of the resonances over this range. To answer the first requirement, aluminum is an appealing material. Aluminum exhibits a broadband plasmonic response, from UV to infrared \cite{gerard15}. It has other advantages, such as being a cheap, abundant and non-critical metal \cite{Graedel:2015jn} and being CMOS compatible \cite{Olson:2014cw}. Aluminum nanostructures have been demonstrated to be interesting plasmonic resonators, either as one-dimensional (nanorods) \cite{knight12,martin2014high} or two-dimensional (nanodisks and nanotriangles) \cite{langhammer08,knight14,campos2017} structures. The increased complexity from 1D to 2D structures yields a richer modal structure. Nanorods sustain only edge modes, which are resonances along the axis of the rod, akin to Fabry-Perot resonances in a one-dimensional optical cavity. In contrast, nanotriangles sustain both edge modes, which are located at the edges of the triangular cavity and correspond to Fabry-Perot resonances along the triangle's sides, and breathing modes, which extend inside the triangle and are somehow analogous to the modes of an acoustic membrane \cite{campos2017}. However, the breathing modes are mostly dark \cite{Schmidt12} so they cannot be excited efficiently by an impinging plane wave. This strongly hinders their use as optical antennas. To obtain a multi-resonant spectrum of \textit{bright modes}, other designs are required.

Inspiration for antenna design can be found in radio-wave science, where fractal antennas have been used for a long time \cite{werner03}. A fractal is, in essence, a self-similar structure that can be created using an iterative process, yielding to multiple copies of an original structure at different scales. Hence, their complexity is controlled by the chosen number of iterations. This is why fractals are interesting for applications requiring compact sizes and/or to optimize the filling of a surface - two key advantages in the design of transparent electrodes \cite{afshin14} or photodetectors \cite{fang16}. To the best of our knowledge, the interest of self-similar structures for optical antennas was first demonstrated in 2015 by Gottheim \textit{et al.} \cite{gottheim2015fractal} using gold Cayley trees. Cayley trees are very simple self-similar structures, yet they exhibit multiple resonances, the number of bright resonances being equal to the number of iterations used to design the tree. Being based on a simple branching geometry, Cayley trees are also much easier to fabricate than other fractal geometries, such as Sierpinsky carpets \cite{volpe11} or Koch flakes \cite{bellido2017self}. 

In this paper, we combine aluminum and fractal plasmonics to design multi-resonant optical antennas. We experimentally demonstrate that a \textit{single} aluminum Cayley tree can sustain multiple resonances over an ultra-broadband range of wavelengths, from the ultraviolet to the mid-infrared (from $\lambda=380$ nm to $\lambda \simeq 10$ $\mu$m). This unprecedented range of tunability is experimentally evidenced using electron energy loss spectroscopy (EELS), a powerful tool allowing to simultaneously obtain a spectrum analogous to an unpolarized optical extinction spectrum \cite{losquin2015unveiling}, and to perform a spatial mapping of the resonances with nanoscale resolution \cite{nelayah2007mapping}. As such, EELS is well-suited to the experimental study of optical antennas and has been applied to the characterization of antennas of various geometries \cite{martin2014high, campos2017, Schmidt12, rossouw2011multipolar}, including fractals \cite{bellido2017self, bicket2020hierarchical, wang2021self}. In order to unveil all the electromagnetic resonances sustained by the Al Cayley trees, it is necessary to probe the mid-IR regime. This might be problematic in EELS, as low-energy resonances are convoluted with the so-called "zero-loss peak", which is associated with the elastic scattering of the electrons \cite{egerton2011electron}. In this work, we use a highly monochromated electron microscope having demonstrated its capability to probe surface plasmons \cite{Mkhitaryan2021} and phonons \cite{Li2021} up to the far-IR. 

\begin{figure}
\centering
\includegraphics[width=0.9\linewidth]{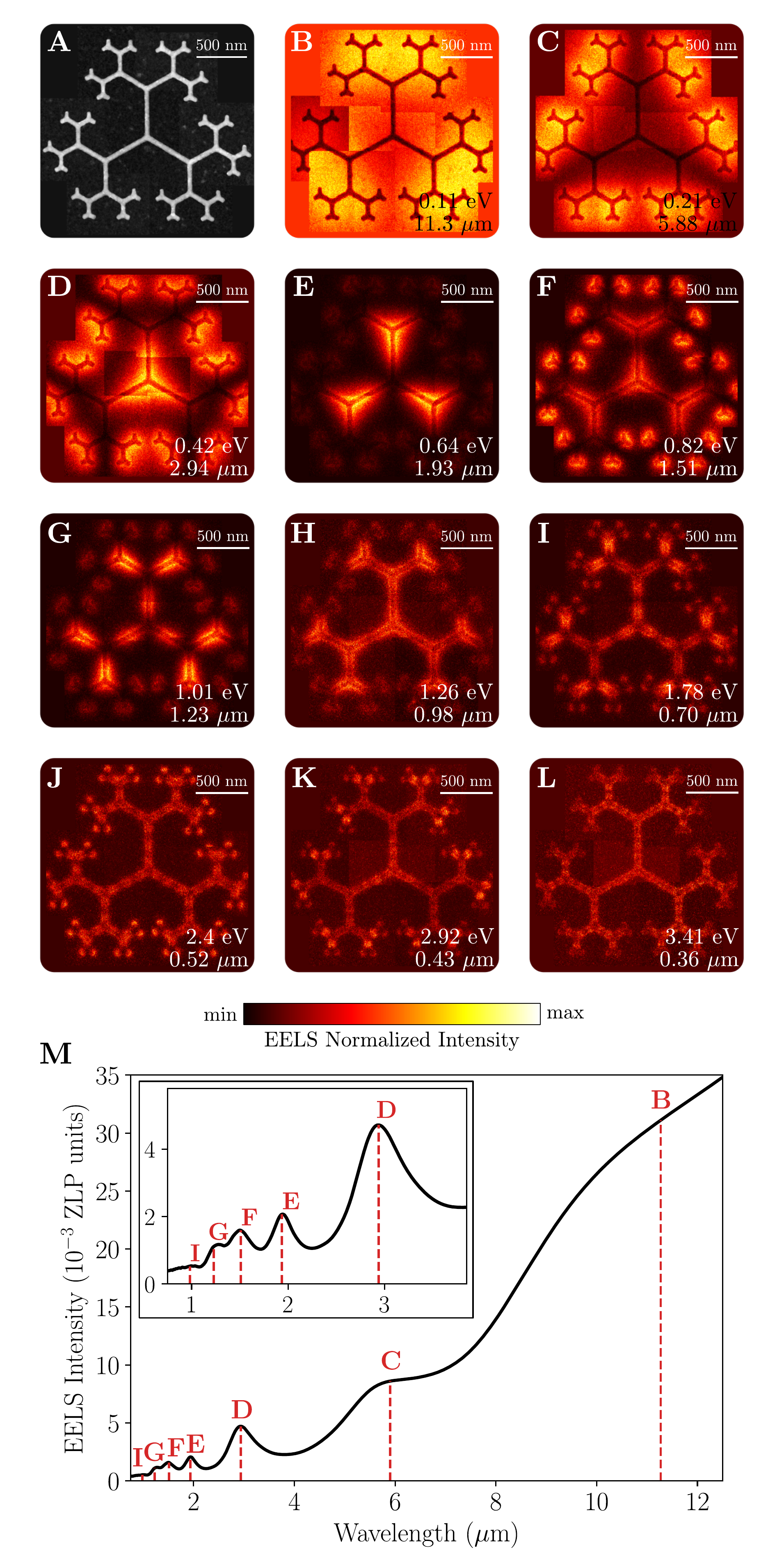}
\caption{Multiple resonances in a single Cayley tree, with generation number $G=5$, aspect ratio $r=0.66$ and initial arm length $L_1=500$ nm. (A) HAADF image showing the entire fractal. (B-L) Energy-filtered EELS images corresponding to different maxima in the EELS spectrum. In order to map the whole structure without image distortion, these images have been reconstituted by stitching several EELS maps with smaller scanning ranges. Please also note that each image has been normalized to its maximum intensity. (M) Averaged EELS spectrum for the Cayley tree. The spectra were normalized by their zero-loss peak intensity then averaged over the whole measured area.}
\label{fig:resonances}
\end{figure}

\section*{Results}
The Cayley trees were designed using the following procedure. Starting from the origin (the tree's \textit{root}), one draws a number of lines (the tree's \textit{branches}) with length $L_1$. In this work, the initial number of branches is set to 3, giving the structure an overall $C_3$ rotational symmetry. This initial structure is the first generation of the tree. Then, at the extremity of each arm, one draws two new branches making an angle of $\pm$60$^{\circ}$ with the original branch and whose length is $L_2=r L_1$, $r$ being the chosen aspect ratio of the tree. This yields the second generation of the Cayley tree. The third generation is obtained by repeating the process: drawing two new branches at the extremity of the first generation branches, whose length is $L_3=r L_2 = r^2 L_1$, and so on for the next generation. This building process is summarized in Fig. S1. A given Cayley tree of $C_3$ rotational symmetry is hence characterized by three numbers: its generation $G$, its initial length $L_1$ and its aspect ratio $r$.

In practice, the Cayley trees were fabricated in aluminum using electron-beam lithography on a substrate compatible with transmission electron microscopy (15-nm-thick Si$_3$N$_4$ membrane). Both the Al thickness and the width of the trees' branches were set to 40-nm (see Methods for details). The Al Cayley trees were then experimentally characterized using EELS in a scanning transmission electron microscope (see Methods). This technique yields simultaneously a high-angle annular dark-field (HAADF) image showing the topography of the structure, and a spectral image \cite{kociak2014mapping}. The EELS signal in the spectral image is almost directly proportional to the $z$-component of the electromagnetic local density of states (EMLDOS) at the position of the electron probe \cite{kociak2008probing}, hence allowing for high spatial resolution mode mapping. Let us also point out that a single EELS spectrum provides information over a broad range of energies, whereas with optical spectroscopy the use of at least two different setups (with different sources, lenses and detectors) would have been required to probe from UV to mid-IR.

\subsection*{Multiple resonances in a single fractal structure}
As a first experimental result, we present the EELS characterization of a large Cayley tree with generation number $G=5$ and initial arm length $L_1=500$ nm. The associated HAADF image is shown in Fig. \ref{fig:resonances}A, evidencing the well-defined structure resulting from the nanofabrication process. In Fig. \ref{fig:resonances}B-L, we present 11 different energy-filtered images corresponding to the 11 maxima experimentally observed in the EELS spectrum of the same Cayley tree. The associated EELS spectrum is shown in Fig. \ref{fig:resonances}M. It has been obtained by normalizing all spectra from an hyperspectral image by their ZLP intensity, before averaging all the normalized spectra. Note that in the case of large structures such as the Cayley tree in Fig. \ref{fig:resonances}, the presented EELS images have been obtained by combining (stitching) several smaller images, in order to limit the image distortion induced by large scanning ranges. No post-treatment has been performed on the reconstituted images, so the stitching limits are clearly visible on the energy-filtered maps (see, e.g., Fig. \ref{fig:resonances}B). 

The energy-filtered EELS images associated with each resonance show a great variety of spatial distribution of the EELS signal inside the antennas, with an increasing complexity when the energy increases. The lowest energy resonance (Fig. \ref{fig:resonances}B) exhibits three spatially extended maxima of EELS intensity located at the Cayley tree's edges. In contrast, the resonance at $2.4$ eV shows a complex distribution with small features at the tips and inside the arms. Altogether, we observe in Fig. \ref{fig:resonances} resonances that are mainly distributed around the tree's extremities - similar to the modes first observed by Gottheim and co-workers in \cite{gottheim2015fractal}. However, we also evidence resonances located around the core of the tree (see, e.g., Fig. \ref{fig:resonances}E), that were not previously reported. Interestingly, comparing Fig. \ref{fig:resonances}B-D shows that some of the resonances seems to exhibit self-similarity in their spatial distribution. This will be discussed in length later. It should be emphasized at this stage that these resonances do not necessarily correspond to well-defined electromagnetic \textit{modes} in the fractal: they can also correspond to a superposition of modes that our limited spectral resolution cannot separate.

This first example of aluminum Cayley tree shows that a single structure can sustain a wealth of resonances over a broad spectral range. Within the limitations of our instrument, resonances are observed from $0.11$ eV (corresponding to a wavelength $\lambda = 10880$ nm) to $3.41$ eV ($\lambda = 364$ nm) -- in other words, from the thermal infrared to the ultraviolet. To the best of our knowledge, such a range of operating wavelengths has never been reported for an optical antenna. In the following, we study how the position and number of resonances can be tuned by changing the geometry of the antenna. Note that in the mid-IR, aluminum behaves as an almost perfect Drude metal, with surface plasmon resonances exhibiting a more "photon-like" behavior - -similarly to what was observed for copper nanostructures in the IR.\cite{Mkhitaryan2021}

\begin{figure*}[t]
\centering
\includegraphics[width=17.8cm]{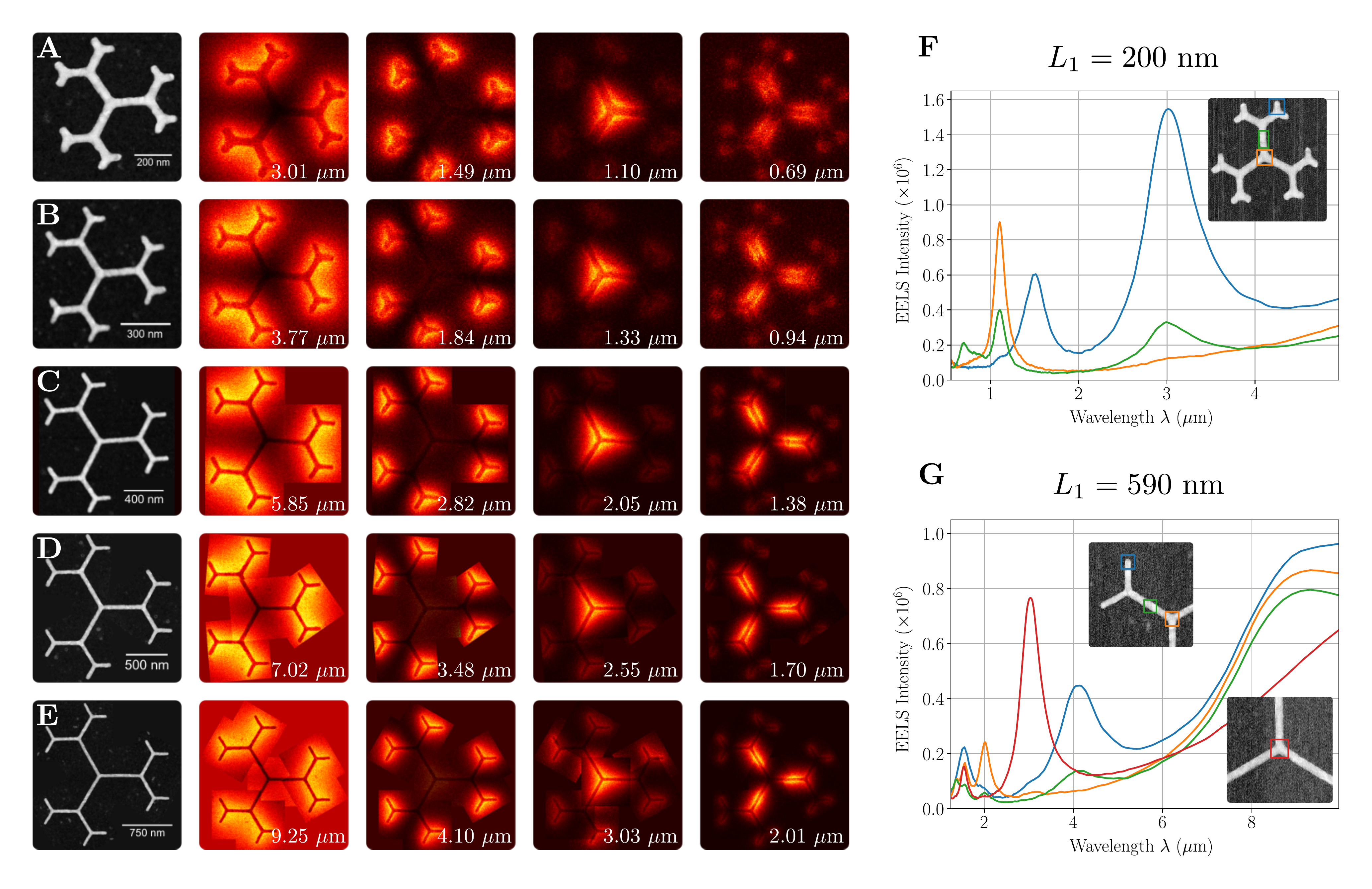}
\caption{Scalable multiple resonances in aluminum Cayley trees. Experimental EELS maps and spectra of Cayley trees exhibiting different initial arm lengths. (A) HAADF image of a Cayley tree with $L_1=200$ nm and energy-filtered EELS maps at four different resonance wavelengths. (B) Same, for $L_1=230$ nm. (B) Same, for $L_1=400$ nm. (D) Same, for $L_1=490$ nm. (E) Same, for $L_1=590$. (F) EELS spectra for the Cayley tree with  $L_1=200$ nm. The spectra were averaged over several pixels, as indicated by the colored boxes in the HAADF image shown as an inset. (G) Same than F, for  $L_1=590$ nm. }
\label{fig:arm}
\end{figure*}

\subsection*{Scalability}
Then, we study the effect of the initial arm length $L_1$. Fig. \ref{fig:arm} shows the EELS maps of four different modes for five different structures. These structures exhibit the same arm width ($w=40$ nm), Al thickness, generation number ($G=3$) and aspect ratio. It is clear from Fig. \ref{fig:arm} that the same modal structures (EELS intensity patterns) are observed in all five structures, but at different wavelengths. For instance, the "dipole-like" resonance from the second column in Fig. \ref{fig:arm}A-E  can be tuned from $\lambda = 3$ $\mu$m to $9.2$ $\mu$m, while the core mode from the forth column in Fig. \ref{fig:arm}A-E is observed from $\lambda = 1.1$ $\mu$m to $3.03$ $\mu$m. This demonstrates that the desired field pattern can be continuously tuned to any wavelength inside the tunability range of aluminium, simply by changing the initial arm length.

Fig. \ref{fig:arm} also present EELS spectra recorded on the smallest ($L_1=200$ nm, Fig. \ref{fig:arm}F) and the largest ($L_1=590$ nm, Fig. \ref{fig:arm}G) antennas. The spectra have been recorded at different locations on the antenna (root, arm center, extremities), as shown by the colored boxes in the inset. As expected, only resonances with a non-zero EELS intensity at the location of the probe are observed in the corresponding spectrum.

\begin{figure*}
\centering
\includegraphics[width=17.8cm]{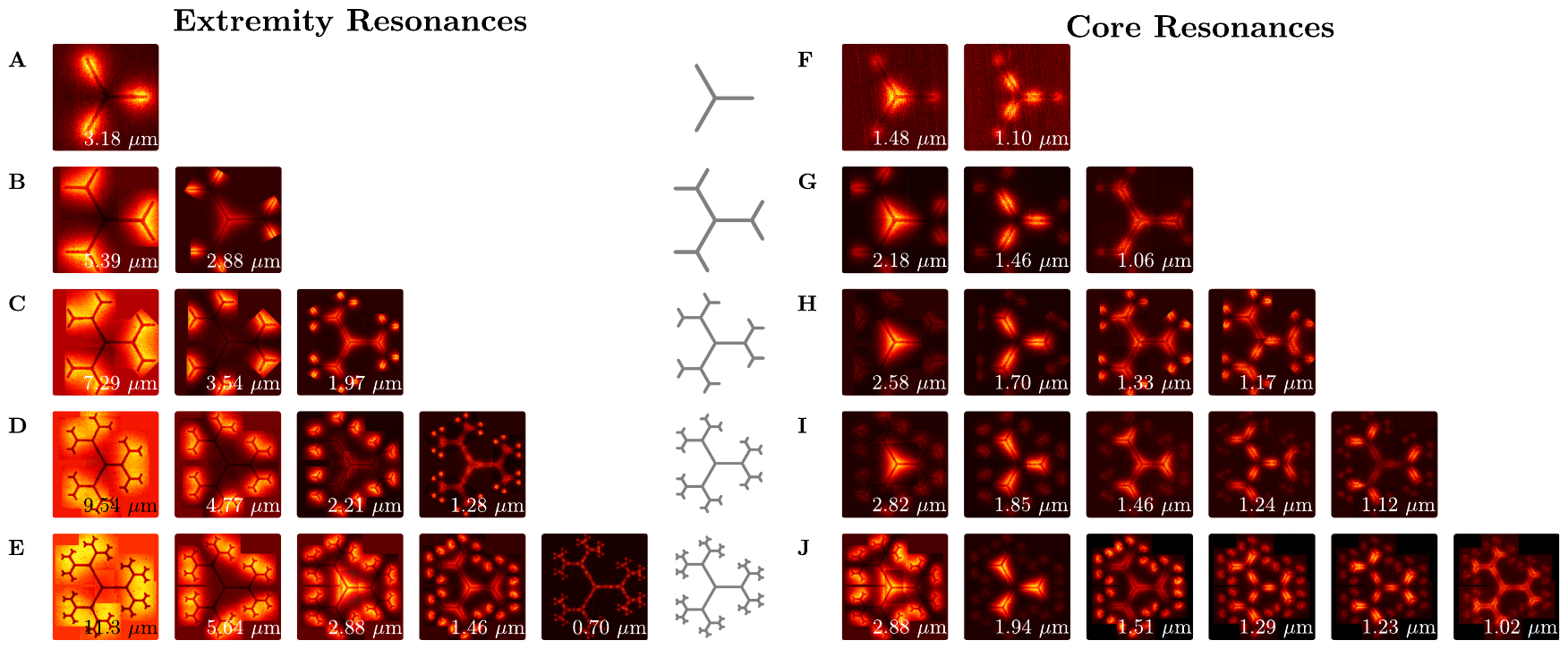}
\caption{Classification of the resonances: energy-filtered EELS maps recorded on five aluminum Cayley trees, all with $L_1=500$ nm and $r=0.625$ but with increasing generation numbers. The resonances are divided between extremity (A-E) and core (F-J) resonances. The generation number is $G=1$ (rows A and F), $G=2$ (rows B and G), $G=3$ (rows C and H), $G=4$ (rows D and I), and $G=5$ (rows E and J).}
\label{fig:extremity}
\end{figure*}

\subsection*{Bright modes}
Now, we focus on the classification of the resonances. Roughly, two main families of modes can be identified. The extremity modes are located on the outer part of the tree, with dipole-like behavior. In contrast, the core modes are located in the center of the tree, mainly on the first generation branches. To go further and to propose a classification, we performed systematic measurements on several Cayley trees of identical initial geometry, with increasing generation number (from $G=1$ to 5). This allowed us to identify similar patterns and to sort out the resonances accordingly. Results are shown in Fig. \ref{fig:extremity}. 

Starting with extremity modes (Fig. \ref{fig:extremity}A-E) we first observe that the number of extremity modes is equal to $G$, i.e. the number of iterations of the self-similar process used to create the fractal-like antenna. Also, it appears clearly that the resonances in the first column in Fig. \ref{fig:extremity}A-E correspond to a "dipole-like" mode, associated with an oscillation involving the whole structure -- similar to the first edge mode observed in nanotriangles \cite{campos2017}. The resemblance with the modes sustained by nanotriangles stops here, as the other resonances are not higher order edge modes, but self-similar copies of the initial extremity mode (compare, for instance, the three extremity modes in Fig. \ref{fig:extremity}C). It should be noted that a self-similar behavior of the mode profile was also observed in Koch flake fractals in \cite{bellido2017self}. This claim of self-similarity is proven in the next section via calculations of the surface charges distribution.

\begin{figure}
\centering
\includegraphics[width=0.9\linewidth]{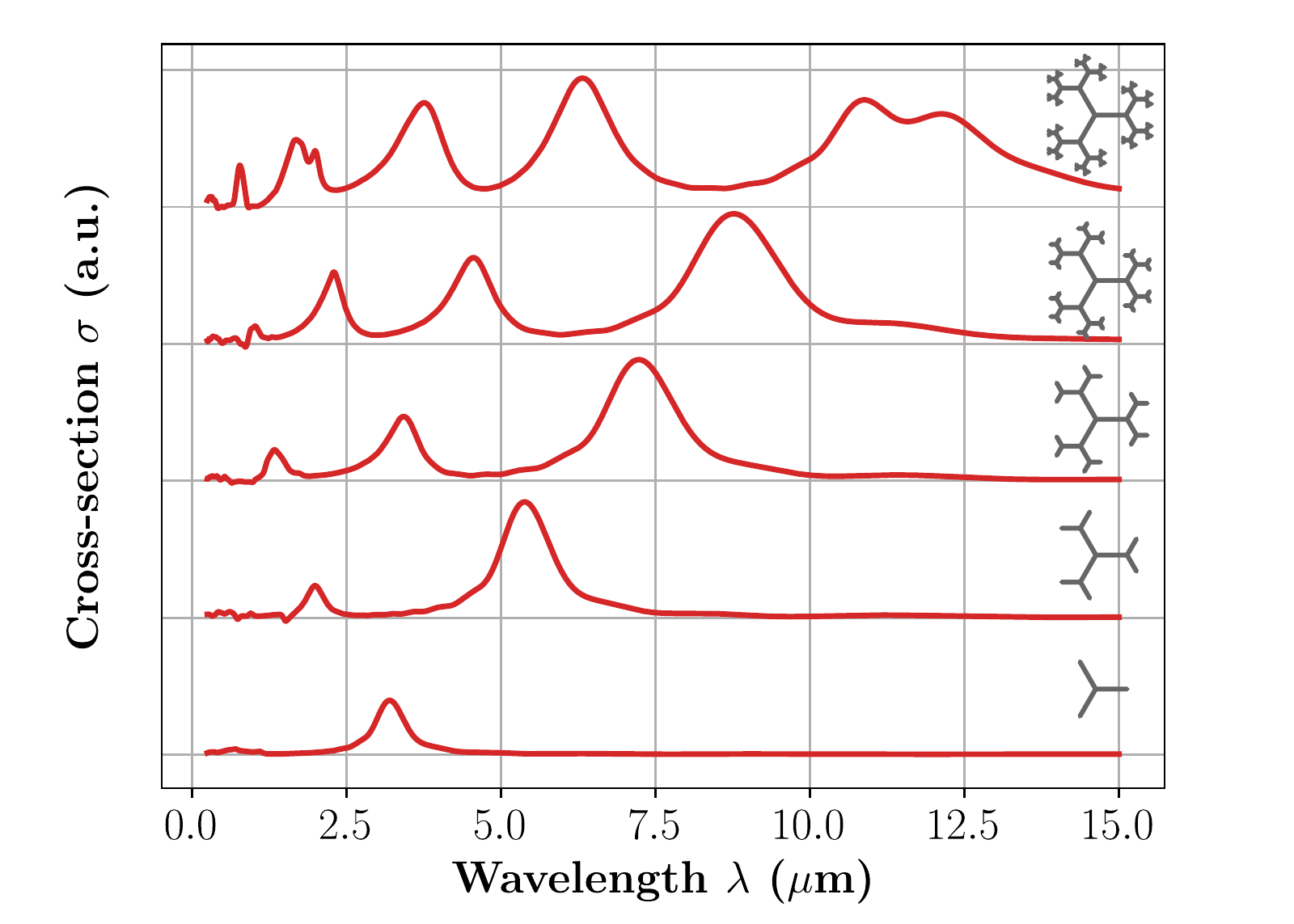}
\caption{FDTD-simulated extinction spectra (extinction cross-section vs. wavelength) for aluminum Cayley trees with different generations, from $G=1$ to $G=5$. The dimensions of the fractals are the same than in Fig. \ref{fig:extremity}. All spectra were plotted using the same linear scale so they can be directly compared, but were vertically offset to ease the reading of the figure.}
\label{fig:fdtd}
\end{figure}

To ascertain the nature of the extremity modes, we performed numerical simulations using the FDTD method (see Methods). In these simulations, the Cayley tree are illuminated with a plane wave. As such, it only reveals the bright modes, in other words modes that are coupled to the radiative continuum. Fig. \ref{fig:fdtd} shows the calculated extinction cross-section of a Cayley tree with initial arm length $L_1=500$ nm, for a number of generations ranging from $G=1$ to 5. The number of resonances in the extinction cross-section is equal to $G$. The link between the number of bright modes and the generation of the fractal was first pointed out in \cite{gottheim2015fractal}. A good agreement between the computed resonance positions and spectral width and the experimental ones is observed. A remarkable exception is the case of the $G=5$ Cayley tree (topmost spectrum in Fig. \ref{fig:fdtd}), as the FDTD spectrum evidences a clear mode splitting for the lower energy mode (first bright mode) as well as for the fourth bright mode (around $\lambda=2$ $\mu$m). This mode splitting is linked to mode hybridization. Due to the high generation number of the fractal, the extremity of the last branches are close, i.e. the distance between neighbouring branches is small when compared with the wavelength. Hence, near-field coupling is observed leading to the apparition of bonding and anti-bonding modes, as in gap antennas. This is demonstrated in the computed charge distributions, evidencing symmetric and anti-symmetric charges across the gap for the anti-bonding and bonding modes, respectively (see Fig. S2 in the Supplementary Information). Interestingly, the long wavelength hybridized mode is not experimentally resolved in EELS, whereas the $\lambda \simeq 2$ $\mu$m is (see Fig. S3 for a direct comparison between the experimental EELS spectrum and the computed one). This is due to the small energy difference between the bonding and anti-bonding modes, which is only $\sim 12$ meV), while the energy resolution of the experimental set-up is around 30 meV (see Methods). In contrast, the splitting of the forth bright mode is experimentally resolved because the energy difference between the two hybridized resonances ($\sim 120$ meV) is larger than the spectral resolution.

\subsection*{Self-similarity of the mode profiles}
\begin{figure}
\centering
\includegraphics[width=.9\linewidth]{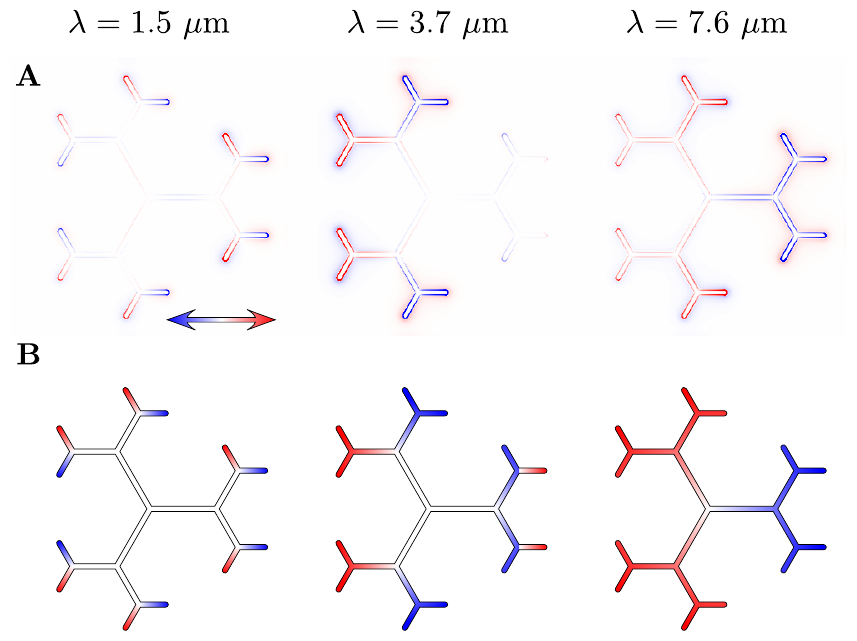}
\caption{FDTD-computed surface electric charges distribution at three different resonance wavelengths for an aluminum Cayley tree with generation number $G=3$, aspect ratio $r=0.625$ and initial arm length $L_1=500$ nm. The excitation is horizontally-polarized (as indicated by the arrow). The plus and minus signs on the figure are a guide for the eye.}
\label{fig:charge}
\end{figure}

In order to demonstrate the claim of self-similarity of the bright modes, plotting the spatial distribution of the surface electrical charges is useful, as surface charges contain a phase information that is missing in the intensity plots. The surface charges distribution can be deduced from the FDTD-computed field distributions (see Methods). Results are shown in Fig. \ref{fig:charge} for the bright modes of a $G=3$ Cayley tree (corresponding to the middle extinction spectrum in Fig. \ref{fig:fdtd}). Here, the surface electric charges have been computed for an horizontally-polarized excitation (see the arrow in Fig. \ref{fig:charge}A). Looking at the surface charges for the lowest energy mode ($\lambda=7.6$ $\mu$m) shows charges of opposite signs on the left and right part of the tree, demonstrating it is a dipole resonance. The next bright mode ($\lambda=3.7$ $\mu$m) shows opposite charges on each second generation branch of the tree, while the high energy mode ($\lambda=1.5$ $\mu$m) shows opposite charges on the third branch of the tree (the plus and minus signs in Fig. \ref{fig:charge} are a guide for the eye). This shows that each mode exhibits a surface charge distribution, which is a down-scaled, self-similar copy of the previous mode.

To further confirm that the self-similarity of the mode profile is induced by the self-similar geometry of the Cayley trees, we performed a control experiment. Instead of a Cayley tree, we designed a "broken line" structure in aluminum (see Fig. S4A). The length and width of this line is the same than of the corresponding Cayley tree, it exhibits similar angles, but its geometry is not self-similar. Results of EELS experiments, as well as FDTD computations, are reported in Fig. S4B-C. The observed mode structure do not exhibit the same wealth of resonances: we only observe resonances similar to those of a bare nanorod of the same total length. This control experiment demonstrates the advantage of using a self-similar structure.

\subsection*{Core modes}
The case of core modes, presented in Fig. \ref{fig:extremity}F-J, is more complex. As previously, similar spatial distributions of the EELS intensity are observed for the different generations, such as the mode located around the tree's root (first column of the core resonances in Fig. \ref{fig:extremity}F-J). However, for $G=5$, the "root mode" appears to be convoluted with another mode located in the outer part of the tree. These two modes cannot be separated with the limited spectral resolution of the instrument. Globally, the EELS maps of the core resonances are much more intricate than the extremity resonances, making it difficult to ascertain if the observed patterns correspond to one or several electromagnetic modes. As these resonances do not appear on the FDTD-calculated extinction spectra, we can also say they are associated with dark modes, i.e. modes that are not efficiently coupled to the radiative continuum. Dark modes can nonetheless be excited by a localized excitation, such as our electron probe, or a nearby emitter.

\section*{Conclusion}
In conclusion, we demonstrated that a single self-similar optical antenna can sustain electromagnetic resonances in a broad spectral range, from the UV to the thermal infrared. This unprecedented range of operation results from both the self-similar design of the antenna and the use of aluminum as a plasmonic metal. Similar antennas made of gold, such as the ones from \cite{gottheim2015fractal}, would not be able to address the UV part of the electromagnetic spectrum. The antennas sustain a wealth of optical resonances, that can be separated into bright "extremity modes" (one bright mode per iteration of the self-similar construction process) and dark "core modes", with more intricate and complex field profiles. The mode spectral position can be easily tuned over the whole operating spectrum, simply by changing the initial arm length. Moreover, it appears that Cayley trees sustains more bright modes than other self-similar geometries, such as Koch fractals \cite{bellido2017self}.

An interesting comparison can be made between the Cayley trees and the canonical nanorod antenna. We compared both the extinction cross-section and the extinction efficiency (i.e., the extinction cross-section divided by the geometrical area of the antenna) for a $G=3$ Cayley tree and a nanorod exhibiting the same total length. Results are shown in Fig. S5 for an unpolarized excitation. If we compare only the dipole resonances, the nanorod is a more efficient antenna (although the Cayley tree present a narrower linewidth). But the spectrum of the self-similar antennas is richer, exhibiting three well-defined resonances. This implies that although nanorods are better \textit{single-mode} antennas, Cayley trees are far superior when multi-mode operation is required.

Broadband, multi-resonant optical antennas could find multiple applications, especially if used as building blocks of a metasurface. An obvious example is non-linear conversion. A non-linear optical medium coupled to a metasurface of such antennas could be pumped in the infrared at a frequency corresponding to a bright mode to generate multiple harmonics with enhanced efficiency. The generated harmonics could be either coupled out of the system (if the harmonic frequency matches another bright mode) or kept in the near-field, if the harmonic frequency corresponds to a dark mode. Interestingly, in nonlinear metasurfaces the selection rules stating which nonlinear processes are allowed are governed both by symmetry of the array and by the symmetry of the individual meta-atoms \cite{li2017nonlinear}. Albeit we only present here $C_3$ symmetry Cayley trees, other rotational symmetries can be created using exactly the same principles, making Cayley trees appealing as meta-atoms for nonlinear metasurfaces. The resonances could also be engineered to match multiple molecular fingerprint bands in infrared spectroscopy, allowing the design of efficient substrates for surface-enhanced infrared absorption (SEIRA). By carefully choosing the position of the antenna's resonances, SEIRA substrates exhibiting electromagnetic selectivity to one or several molecular species could be designed. THz generation \cite{mcdonnell2021functional} and photodetection \cite{fang16} are other possible applications.

\section*{Materials and methods}
\subsection*{Sample fabrication}
The antennas were fabricated using electron beam lithography (EBL) in a FEG SEM system (eLine, Raith). First, a 150 nm-thick layer of poly(methyl methacrylate) resist was spin-coated on a scanning TEM-EELS compatible substrate. The latter consists in arrays of 15 nm-thick Si$_3$N$_4$ square membranes engraved in a small silicon wafer with a 3 mm diameter (from Ted Pella, Inc). The resist was subsequently insulated by the electron beam using the EBL system. The imprinted patterns were then developed during 60 s in a 1:3 MIBK:IPA solution at room temperature. Then, a 40 nm-thick layer of Al was deposited on the sample using thermal evaporation (ME300, Plassys). Finally, lift-off has been accomplished by immersing the sample in acetone unveiling the Al nanoantennas on the membranes. 

\subsection*{EELS experiments and data analysis}
EELS spectroscopy and mapping were performed with a monochromated NION Hermes200 microscope fitted with an IRIS spectrometer. The acceleration voltage was 100 keV and the energy resolution was in the range 24-30 meV. Typical spectral images were 200 per 200 pixels and the size adjusted to the given structure. All spectra have been aligned a posteriori based on the zero-loss peak position. For the larger fractal antennas, several images were recorded and subsequently stitched. For all the energy-filtered maps presented in this paper, the color scale has been normalized in order to maximize the visibility of the image, so the intensity of the EELS signal cannot be directly compared from one map to another.
 
\subsection*{Numerical simulations}
Simulations were performed using a commercial code (Lumerical FDTD Solutions). A constant mesh size of 5 nm was set to define precisely each antenna, while a non-uniform mesh was used outside the antennas. We used perfectly absorbing layers (PMLs) on the six sides of the computation box, as well as a symmetric boundary condition along the excitation axis, aligned with one arm of the simulated Cayley tree. All simulations were performed for 40-nm thick Al Cayley trees deposited on top of a 15-nm thick Si$_3$N$_4$ membrane, reproducing the experimental configuration. Moreover, we used in computations the experimental sizes as directly measured on the HAADF images. In particular, we observed that the fabricated width of the tree's branches was slightly different from the 40-nm setting. In practice we measured widths ranging from $42\pm 3$ nm to $48\pm 4$nm (depending on the tree's generation number), resulting in a non-negligible effect on the computed position of the resonances. The refractive indices were taken directly from the software's library of materials.
The spatial distribution of the electrical surface charges is directly computed from the electric field components via the equation $\rho / \varepsilon= \overrightarrow{\nabla} \cdot \overrightarrow{E}$, where $\varepsilon$ is the dielectric permittivity of the surrounding medium.

\begin{acknowledgments}
TS and DK acknowledges support from the R\'egion Grand Est. Samples were realized on the Nanomat platform (www.nanomat.eu). This work has been done within the framework of the Graduate School NANO-PHOT (grant ANR-18-EURE-0013). This work received support from the National Agency for Research under grant QUENOT (ANR-20-CE30-0033) and the program of future investment TEMPOS-CHROMATEM (ANR-10-EQPX-50). This project also received funding from the European Union's Horizon 2020 Research and Innovation Program under grants 823717 (ESTEEM3) and 101017720 (EBEAM). The authors acknowledge financial support from the CNRS-CEA "METSA" French network (FR CNRS 3507) on the plateform LPS-STEM.
\end{acknowledgments}

\bibliography{PNAS-fractal-main}

\end{document}